\pdfoutput=1

\documentclass[10pt,amsmath,amssymb,aps,pra,twocolumn,longbibliography,floatfix,superscriptaddress,nofootinbib]{revtex4-1}

\usepackage[utf8]{inputenc}
\usepackage[T1]{fontenc}

\usepackage{dsfont}
\usepackage{times}
\usepackage{txfonts}

\usepackage{physics}
\usepackage{mathtools}
\usepackage{mleftright}\mleftright
\usepackage{float}
\usepackage{enumitem}
\usepackage{algorithm}
\usepackage{algpseudocode}
\floatstyle{boxed}
\newfloat{algorithm}{t}{lop}
\floatname{algorithm}{Algorithm} 

\usepackage{thmtools} 
\usepackage{thm-restate}

\usepackage{multirow}
\usepackage{diagbox}
\usepackage{soul}

\floatstyle{boxed}

\newfloat{algorithm}{t}{lop}
\floatname{algorithm}{Algorithm} 

\newfloat{metaalgorithm}{t}{lop}
\floatname{metaalgorithm}{Meta-Algorithm}

\newtheorem{problem}{Problem} 
\newtheorem{lemma}{Lemma}
\newtheorem{corollary}[lemma]{Corollary}
\newtheorem{definition}[lemma]{Definition}

\newenvironment{proofsketch}{%
\par
\noindent
\textit{Proof (sketch):}
\noindent}
{\hfill$\square$
\smallskip}

\newenvironment{proof}{%
\par
\noindent
\textit{Proof:}
\noindent}
{\hfill$\square$
\smallskip}

\mathchardef\mhyphen="2D

\DeclareMathOperator*{\Ex}{\mathbf{E}}
\let\Pr\relax
\DeclareMathOperator*{\Pr}{\mathbf{Pr}}

\newcommand{\lr}[1]{\left(#1\right)}
\newcommand{\lrq}[1]{\left[#1\right]}
\newcommand{\lrb}[1]{\left\{#1\right\}}

\newcommand{\mc}[1]{\mathcal{#1}}
\newcommand{\mb}[1]{\mathbb{#1}}
\newcommand{\mbf}[1]{\mathbf{#1}}

\renewcommand{\eval}{\mathtt{Eval}}

\newcommand{\mq}{\mathtt{MQ}}

\newcommand{\samp}{\mathtt{Samp}}
\newcommand{\stat}{\mathtt{Stat}}
\newcommand{\gen}{\mathtt{Gen}}

\newcommand{\eps}{\varepsilon}
\renewcommand{\epsilon}{\varepsilon}

\def\01{\{0,1\}}

\newcommand{\tv}{\mathrm{TV}}

\newcommand{\D}{\mc D}
\newcommand{\A}{\mc A}
\newcommand{\G}{\mc G}
\newcommand{\B}{\mc B}

\newcommand{\Cl}{\mathrm{Cl}}

\newcommand{\poly}{{\rm{poly}}}

\usepackage{dcolumn}
\usepackage{bm}

\usepackage{graphicx}
\usepackage{xcolor}

\usepackage[pagebackref,final,colorlinks,allcolors = blue]{hyperref}
\hypersetup{pdftitle = {A single $T$-gate makes distribution learning hard},
    pdfauthor = {
        Marcel Hinsche, 
        Marios Ioannou, 
        Alexander Nietner, 
        Jonas Haferkamp, 
        Yihui Quek, 
        Dominik Hangleiter, 
        Jean-Pierre Seifert, 
        Jens Eisert, 
        Ryan Sweke},
    pdfsubject = {
        Machine learning, 
        quantum computation},
    pdfkeywords = {
        quantum machine learning, 
        PAC learning, 
        quantum advantage, 
        Born machine,
        quantum circuit, 
        Clifford circuits,
        statistical query model, 
        generative modelling, 
         }
      }

\usepackage{cleveref}

\begin{document}

\title{A single $T$-gate makes distribution learning hard}

\newcommand{\fu}{Dahlem Center for Complex Quantum Systems, Freie Universit\"{a}t Berlin, 
14195 Berlin, Germany}

\newcommand{\QuICS}{Joint Center for Quantum Information and Computer Science (QuICS), University of Maryland \& NIST, College Park, MD 20742, USA}

\newcommand{\HZB}{Helmholtz-Zentrum Berlin f{\"u}r Materialien und Energie, 14109 Berlin, Germany}

\newcommand{\HHI}{Fraunhofer Heinrich Hertz Institute, 10587 Berlin, Germany}

\newcommand{\tu}{Department of Electrical Engineering and Computer Science, TU Berlin, 10587 Berlin, Germany}

\author{M. Hinsche} 
\thanks{M.~H.~,~M.~I.~,~A.~N.~and R.~S.~have contributed equally.\\
Corresponding authors: m.hinsche@fu-berlin.de, marios.ioannou@fu-berlin.de, a.nietner@fu-berlin.de, rsweke@gmail.com}
\affiliation{\fu}

\author{M. Ioannou}
\thanks{M.~H.~,~M.~I.~,~A.~N.~and R.~S.~have contributed equally.\\
Corresponding authors: m.hinsche@fu-berlin.de, marios.ioannou@fu-berlin.de, a.nietner@fu-berlin.de, rsweke@gmail.com}
\affiliation{\fu}

\author{A. Nietner}
\thanks{M.~H.~,~M.~I.~,~A.~N.~and R.~S.~have contributed equally.\\
Corresponding authors: m.hinsche@fu-berlin.de, marios.ioannou@fu-berlin.de, a.nietner@fu-berlin.de, rsweke@gmail.com}
\affiliation{\fu}

\author{J. Haferkamp}
\affiliation{\fu}

\author{Y.~Quek}
\affiliation{\fu}

\author{D. Hangleiter}
\affiliation{\QuICS}

\author{J.-P. Seifert}
\affiliation{\tu}

\author{J. Eisert}
\affiliation{\fu}
\affiliation{\HZB}
\affiliation{\HHI}

\author{R. Sweke}
\thanks{M.~H.~,~M.~I.~,~A.~N.~and R.~S.~have contributed equally.\\
Corresponding authors: m.hinsche@fu-berlin.de, marios.ioannou@fu-berlin.de, a.nietner@fu-berlin.de, rsweke@gmail.com}
\affiliation{\fu}

\date{\today}

\begin{abstract}
\noindent The task of learning a probability distribution from samples is ubiquitous across the natural sciences.
The output distributions of local quantum circuits form a particularly interesting class of distributions, of key importance both to quantum advantage proposals and a variety of quantum machine learning algorithms. In this work, we provide an extensive characterization of the learnability of the output distributions of local quantum circuits. Our first result yields insight into the relationship between the efficient learnability and the efficient simulatability of these distributions. Specifically, we prove that the density modelling problem associated with Clifford circuits can be efficiently solved, while for depth $d=n^{\Omega(1)}$ circuits the injection of a single $T$-gate into the circuit renders this problem hard.
This result shows that efficient simulatability does not imply efficient learnability.
Our second set of results provides insight into the potential and limitations of quantum generative modelling algorithms. We first show that the generative modelling problem associated with depth $d=n^{\Omega(1)}$ local quantum circuits is hard for \textit{any} learning algorithm, classical or quantum. As a consequence, one \textit{cannot} use a quantum algorithm to gain a practical advantage for this task. We then show that, for a wide variety of the most practically relevant learning algorithms -- including hybrid-quantum classical algorithms --  even the generative modelling problem associated with depth $d=\omega(\log(n))$ Clifford circuits is hard. This result places limitations on the applicability of near-term hybrid quantum-classical generative modelling algorithms. 
\end{abstract}

\maketitle

\noindent Deep generative models have recently empowered many impressive scientific feats, ranging from predicting protein structure to atomic accuracy \cite{alphafold21} to achieving human-level language comprehension  \cite{Chinchilla22}. 
Consequently, there has been much interest in architecture and algorithm development for probabilistic modelling. Ideally one would like to obtain a rigorous theoretical understanding of these emerging state-of-the-art models, which requires a suitable theoretical framework. Such a framework is provided by the problem of \textit{distribution learning}: 
Given samples from an unknown distribution, output some suitable representation of that distribution. Significant effort has been devoted to characterizing the complexity of learning various classes of structured distributions \cite{canonne2020short,diakonikolas2016learning, Kearns:1994:LDD:195058.195155}, including mixture models~\cite{diakonikolasStatisticalQueryLower2017,   chanLearningMixturesStructured2013}, output distributions of restricted Boolean circuits~\cite{Kearns:1994:LDD:195058.195155,deLearningSatisfyingAssignments2015a} and Poisson binomial distributions~\cite{daskalakisLearningPoissonBinomial2012}. However, these classes of distributions are still somewhat removed from those of most interest to machine learning practitioners, such as those governing movements in the stock market, or the outputs of deep generative models.

Simultaneously, the last years have witnessed significant interest in the potential of exploiting quantum devices for machine learning tasks~\cite{biamonte2017quantum,RevModPhys.91.045002,PhysRevLett.116.250501}. Of particular interest are hybrid quantum-classical schemes, in which parameterized quantum circuits are used as a model class, whose parameters are optimized via classical algorithms~\cite{bharti2021noisy,benedetti2019parameterized}. In the context of generative modelling, the output distributions of quantum circuits are a particularly natural model class, referred to as \textit{quantum circuit Born machines} (QCBMs) \cite{Benedetti_2019,liuDifferentiableLearningQuantum2018}.
In particular, it is known that this model class is expressive enough to contain many probabilistic graphical models \cite{glasser2019expressive,quantumtensors}, while not being classically simulatable \cite{Bremner_2010,boixo_characterizing_2016,SamplingReview2022}. 
These facts, along with a growing body of numerical experiments \cite{coyleBornSupremacyQuantum2020,generativegeneral,rudolph2020generation,niu2020learnability}, suggest that hybrid quantum-classical algorithms using QCBMs as a model class may offer concrete advantages over state-of-the-art classical generative modelling techniques. However, to date, there are no rigorous results on the learnability of this model class which support this intuition.

In order to address this, we provide in this letter a comprehensive study of the learnability of the output distributions of local quantum circuits -- i.e., QCBMs.
This allows us to resolve a variety of open questions.
Firstly, we provide two hardness results for the generative modelling problem associated with these distributions. The first shows that the output distributions of $n$ qubit quantum circuits of depth $n^{\Omega(1)}$ 
are not efficiently learnable by any learning algorithm with access to samples from the unknown distribution.
The second shows that the output distributions of quantum circuits of depth $\omega(\log(n))$ are not efficiently learnable by algorithms which use only statistical averages with respect to the unknown distribution.
Most practically relevant algorithms are indeed of this type.
To date, the output distributions of local quantum circuits are 
considered the most promising candidate for demonstrating a rigorous complexity theoretic separation between the power of QCBM-based hybrid quantum-classical algorithms and purely classical generative modelling techniques. However, our hardness results show that this is not possible, and, therefore, place strong limitations on the advantages one might hope to achieve in this setting with near-term quantum devices.

\begin{figure*}
\includegraphics[scale=.55]{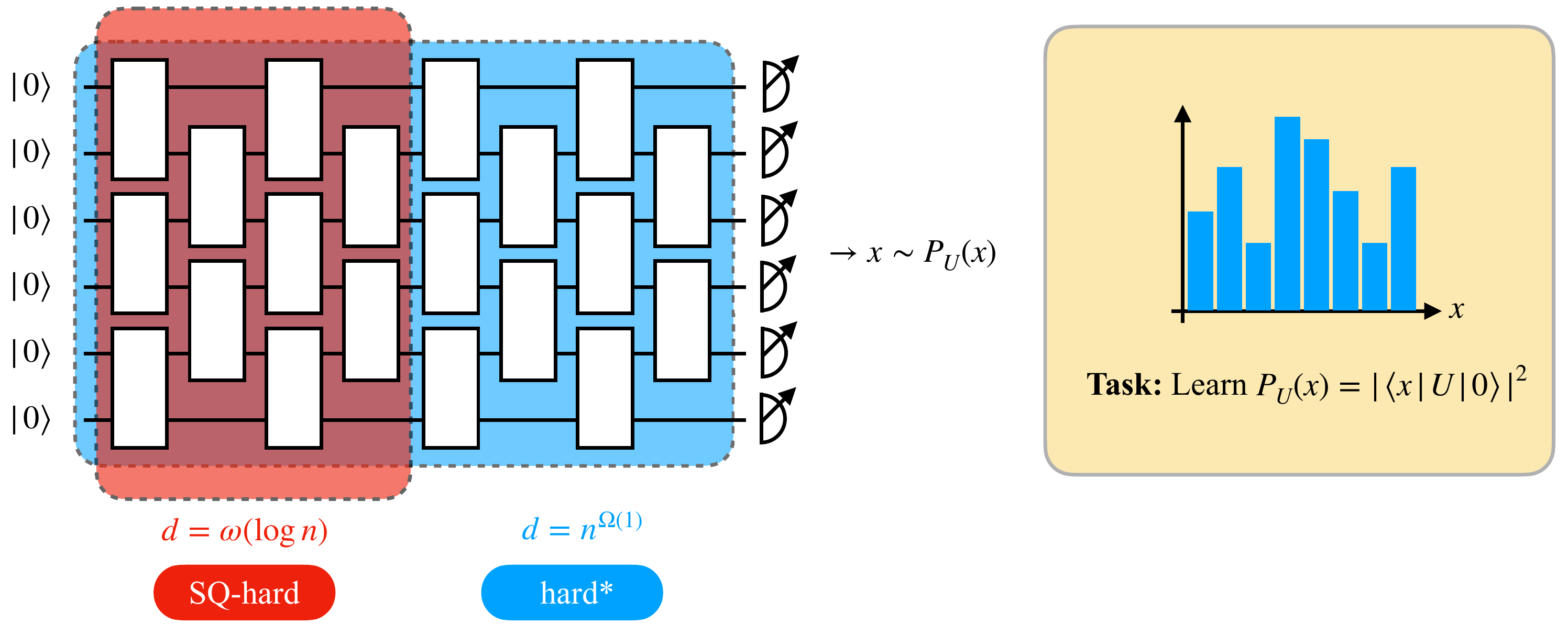}
\caption{How hard is the task of generator- or evaluator-learning the output distributions of local quantum circuits on $n$ qubits of depth $d$? In accord with the intuition that deeper circuits generate more complex distributions, our answer depends on how $d$ scales with $n$. We find that for $d= \omega(\log n)$, even the output distributions of Clifford circuits are not efficiently learnable when given {\em statistical query} access to $P_U$ (Theorem~\ref{SQhard}).  When given the {\em sample} access, the output distributions of generic local quantum circuits cease to be efficiently learnable at linear depths $d = n^{\Omega(1)}$ and beyond, up to standard cryptographic assumptions (Corollary~\ref{cor:eval_linear} and Theorem~\ref{PRFhard}). }
\end{figure*}

Secondly, we show clearly that, within the context of distribution learning, classical simulatability of a class of quantum circuits \textit{does not} imply efficient learnability. This is in strong contrast to existing conjectures and known results in other related settings
\cite{coyleBornSupremacyQuantum2020, Rochhetto,laiLearningQuantumCircuits2021a, 2019YoganathanCondition}.
To do this, we prove that the output distributions of Clifford circuits are efficiently learnable, while the addition of a single $T$-gate to the circuit renders the learning problem hard.
As such, while the complexity of the classical simulation scales with the number of $T$-gates, we find that the addition of a single $T$-gate induces a striking complexity transition in the corresponding distribution learning problem.

\smallskip
{\noindent\it \textbf{Setting.}} --- 
In this Letter, we are concerned with learning distributions promised to be from a \emph{distribution class} $\mathcal{D}$.
In particular, we are interested in the properties of learning algorithms that solve the following problem
\footnote{$\mathrm{TV}$ denotes here the total
variation distance between two probability distributions, see also the appendix.}:

\begin{problem}[Distribution learning]\label{prob:distribution-learning} Given a distribution class 
$\mathcal D$, samples from an unknown distribution $P\in\mathcal{D}$,
and $\epsilon,\delta \in (0,1)$, output with probability at least $1-\delta$, a
representation of a distribution $Q$ satisfying $\mathrm{TV}(P,Q)\leq \epsilon$.
\end{problem}

\noindent We will be concerned with two types of 
\emph{representations}, namely \emph{generators} and \emph{evaluators}:
\begin{itemize}
    \item An evaluator for a distribution $Q$ is a computationally efficient algorithm which, when given some $x$, outputs the probability $Q(x)$.
    \item A generator for a distribution $Q$ is a computationally efficient algorithm for generating samples from $Q$. 
\end{itemize}
We note that the problem of distribution learning with respect to an evaluator is often referred to as \textit{density modelling}, while the problem of learning with respect to a generator is often referred to as \textit{generative modelling}.
Additionally, we stress that in the case of generative modelling it is \textit{not} sufficient for the learning algorithm to store and later reproduce the samples it received during the learning phase, or to output a larger but still bounded set of samples \cite{pmlr-v119-axelrod20a}. Indeed, the learning algorithm is required to output another algorithm -- a generator -- which can output as many as samples as desired, from a distribution which is close in total variation distance to the unknown target distribution.

We are concerned here exclusively with discrete distributions over $\{0,1\}^n$, and denote the set of all such distributions by $\mathcal{D}_n$.
Given some $\mathcal{D}\subseteq\mathcal{D}_n$, we say that an algorithm is a computationally (sample) efficient algorithm for learning $\mathcal{D}$ with respect to a particular representation (either generators or evaluators) if it solves the above problem for all $P\in\mathcal{D}$, using $O(\mathrm{poly}(n,1/\epsilon,1/\delta))$ computational time (samples). If there exists a computationally efficient learning algorithm for $\mathcal{D}$ with respect to a particular representation, then we say that $\mathcal{D}$ is efficiently learnable with respect to that representation. If there does not exist a computationally efficient learning algorithm for some class $\mathcal{D}$ with respect to a particular representation, then we say that $\mathcal{D}$ is hard to learn with respect to that representation. 

Our particular focus in this work is on the output distributions of quantum circuits. More specifically, to any unitary $U$ we have the associated probability distribution $P_U$, with probabilities
\begin{equation}
P_U(x):=\left|\,\langle x | U|0^{\otimes n}\rangle\,\right|^2.
\end{equation}
We then consider sets of distributions obtained from all unitaries generated by quantum circuits of a specific depth, with gates from a specific gate set. Unless otherwise specified, we consider one-dimensional circuits consisting only of nearest-neighbour gates, which for convenience we refer to as \emph{local} quantum circuits. We are particularly interested in how the complexity of learning depends on both the gate set, and the circuit depth. We note that our results generalize and extend seminal work on learning the output distributions of classical circuits~\cite{Kearns:1994:LDD:195058.195155}.

\smallskip
{\noindent\it \textbf{Learning Clifford distributions.}} --- We start by studying the learnability of the output distributions of Clifford circuits. Our primary motivation for doing so is to better understand the relation between the complexity of classical simulation of quantum circuits and their learnability: It is well-known that by
virtue of the Gottesman-Knill theorem, Clifford circuits can be
efficiently classically simulated \cite{gottesman1998heisenberg, Aaronson_2004}.
Similarly, it has been found previously that the algebraic structure of the Clifford group also facilitates efficient learning of an unknown stabilizer state \cite{montanaro2017learning} or Clifford circuit \cite{laiLearningQuantumCircuits2021a} from few copies of the unknown quantum state. 
Furthermore, stabilizer states have been found to be efficiently PAC-learnable \cite{Rochhetto,gollakota2021hardness} in Aaronson's framework for PAC-learning quantum states \cite{Aaronson_2007}. 
In this setting, Ref.~\cite{2019YoganathanCondition} finds a sufficient condition under which the complexity of simulatibility and learnability are aligned. 
Here, we ask whether the alignment in the complexity of classical simulation and learning holds also in the distribution learning setting.
Indeed, when studying Clifford circuits, we find that our learning model is no exception. 

\begin{restatable}{theorem}{CliffordLearning}
\label{Clifford_learning}
The set $\D_\Cl$ of Clifford circuit distributions, for any depth, is efficiently learnable with respect to  generators and evaluators.
\end{restatable}

\begin{proofsketch} Clifford circuit output distributions are uniform over affine subspaces of the finite $n$ dimensional vector space $\mathbb{F}_2^n$. Hence using Gaussian elimination on $O(n)$ samples recovers the correct affine subspace, and from this the correct distribution representation, with success rate $1-\exp(-\Omega(n))$.
\end{proofsketch}

{\noindent\it \textbf{Hardness of learning Clifford$+T$-distributions.}} --- 
Next, we ask whether this alignment of complexity extends even to slightly non-Clifford circuits. In particular, on the simulation side, the run-time of the best-known classical algorithms for simulating $T$-enriched Clifford circuits will grow exponentially with the number of $T$ gates \cite{PhysRevLett.116.250501,PhysRevLett.115.070501,Hakop2020, Bravyi2019simulationofquantum}. 
On the learning side, a first result for learning output states of unknown Clifford+$T$ circuits, from copies of the unknown state, has been obtained in Ref.~\cite{laiLearningQuantumCircuits2021a}. They also find an exponential scaling in the number of $T$ gates provided all $T$ gates are applied in a single layer.

Let us now return to the distribution learning setting. We consider the class of output distributions arising from $T$-enriched Clifford circuits. 
The following result relies on the \emph{learning parities with noise} (LPN) assumption. It posits that there does not exist an efficient algorithm, quantum or classical, for learning from classical samples the class of Boolean parity functions under the uniform distribution when subject to any constant-rate random classification noise. We note that this is a canonical assumption for many cryptographic schemes \cite{regevOnLattices2009, pietrzakCryptography2012}.

\begin{restatable}{theorem}{CliffordPlusT}
\label{thm:Clifford_T_eval_hardness}
Under the LPN assumption, the output distributions of local Clifford circuits of depth $d=n^{\Omega (1)}$ enriched with a single $T$-gate are not efficiently learnable with respect to an evaluator. 
\end{restatable}

\begin{figure}
\includegraphics[scale=0.5]{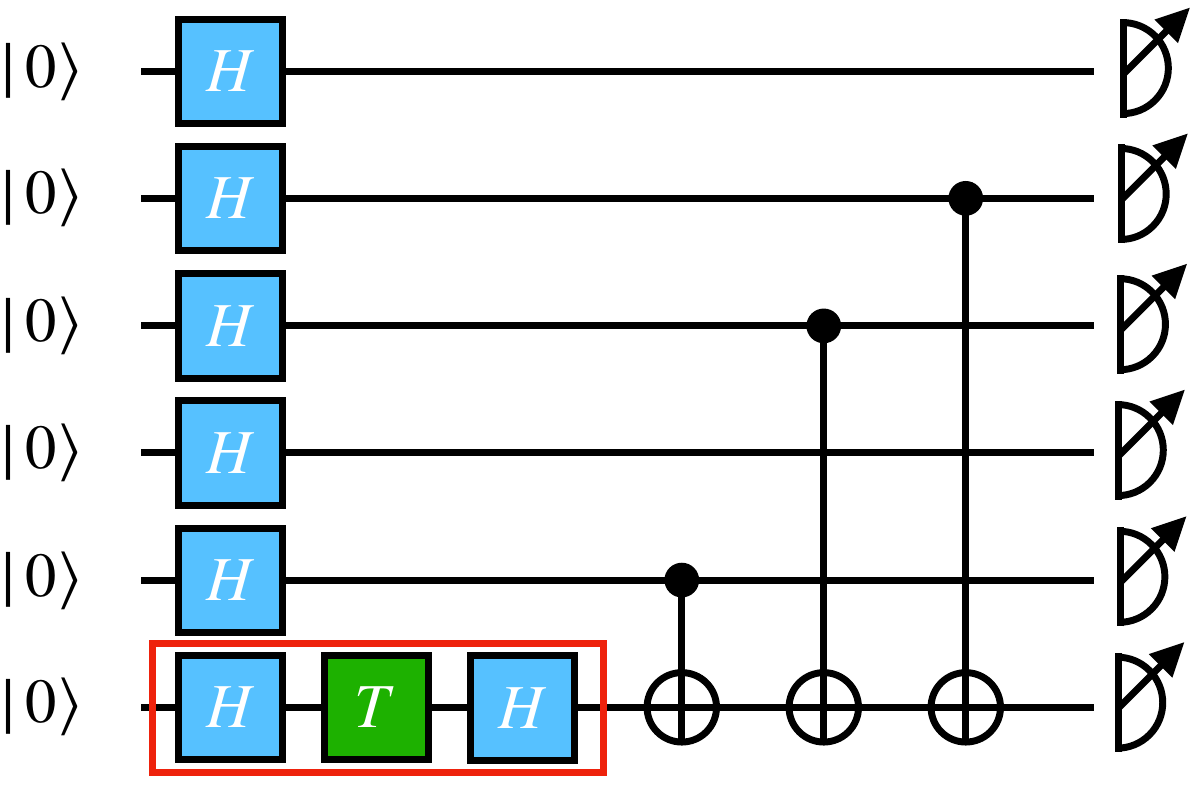}
\caption{Example of a circuit used in the proof of Theorem \ref{thm:Clifford_T_eval_hardness}. Without the red box, samples from this circuit are of the form $(x,f(x))$ where $x$ is uniformly random and $f$ is the parity function supported on bits $2,3,5$. With the red box, the samples are of the form $(x,y)$ where $y=f(x)$ with probability $1-\eta$ and $y = \lnot f(x)$ with probability $\eta$.}
\label{fig:parity_with_noise}
\end{figure}
\begin{proofsketch}
Ref.~\cite{Kearns:1994:LDD:195058.195155} gives a class of distributions such that LPN reduces to evaluator-learning this class. 
Specifically, for each parity function, there is a corresponding distribution. Each such distribution can be realized as the output distribution of Clifford circuit enriched with a single $T$ gate (see e.g. Fig. \ref{fig:parity_with_noise}). We obtain the stated depth dependence by recompiling the circuit into local gates and using a rescaling argument to trade circuit-depth for learning complexity.
\end{proofsketch}

We note that a similar hardness result based on the LPN assumption can be obtained for output distributions of Clifford circuits subject to single-qubit depolarizing noise. The key insight underlying the proof of \Cref{thm:Clifford_T_eval_hardness} is that the LPN noise can be realized by a single $T$ gate. Moreover, it can be seen that, if one relaxes the nearest-neighbour requirement on the Clifford gates, i.e., allowing instead for arbitrary connectivity between qubits, then one obtains the above hardness result in Theorem \ref{thm:Clifford_T_eval_hardness} already for depth $d=\Omega(1)$. 

The sharp transition in complexity between Theorem \ref{Clifford_learning} and Theorem \ref{thm:Clifford_T_eval_hardness} stands in interesting contrast to the smooth increase in the complexity of classically simulating $T$-enriched Clifford circuits: In particular, while $T$-enriched Clifford circuits can be simulated efficiently for up to $O(\log n)$ many $T$ gates \cite{Bravyi2019simulationofquantum}, a single $T$ gate is enough to make distribution-learning with an evaluator at least as hard as LPN.

The class of $T$-enriched local Clifford circuits is a subclass of the class of all local quantum circuits. Hence, the conditional hardness result of Theorem \ref{thm:Clifford_T_eval_hardness} also applies to this more general class:

\begin{corollary}\label{cor:eval_linear} 
Under the LPN assumption, the output distributions of local quantum circuits of depth $d=n^{\Omega(1)}$ are not efficiently learnable with respect to an evaluator.
\end{corollary}

\smallskip
{\noindent\it \textbf{Hardness of learning  generators.}} --- In the previous sections we have seen how adding a single $T$ gate can make the task of learning an evaluator for Clifford distributions at least as hard as LPN. This leaves open the question of the complexity of learning the output distributions of non-Clifford circuits with respect to a generator. 
As discussed in the introduction, the complexity of generator learning is interesting not only from a purely theoretical standpoint. It also allows us to gain insight into the potential of quantum generative models (QCBMs). 

In Ref.~\cite{Kearns:1994:LDD:195058.195155}, it has been shown that the output distributions of polynomially sized classical circuits are not efficiently classically learnable with respect to a generator. In this section, we establish an analogous result for the output distributions of quantum circuits by adapting the proof strategy of Ref.~\cite{Kearns:1994:LDD:195058.195155}. Our result applies to both quantum and classical learning algorithms. In particular, we show that one can embed \emph{pseudorandom functions} (PRFs) into the output distributions of local quantum circuits. In order to establish hardness for quantum learning algorithms, we use ``standard-secure" PRFs -- i.e., PRFs secure against quantum adversaries with classical membership queries~\cite{zhandry2012construct}.

\begin{restatable}{theorem}{PRFhard}
\label{PRFhard}
Assuming the existence of classical-secure (standard-secure) pseudorandom functions, there is no efficient classical (quantum) algorithm for learning the output distributions of depth $d=n^{\Omega(1)}$ local quantum circuits, with gates from any universal gate set.
\end{restatable}

\begin{proofsketch} Instantiating the proof of Theorem 17 in Ref.~\cite{Kearns:1994:LDD:195058.195155} with a standard-secure PRF yields the following: the output distributions of polynomially sized classical circuits are not efficiently generator learnable, even by quantum learning algorithms. Polynomially sized classical circuits can be realized by polynomially sized local quantum circuits. Therefore, the output distributions of polynomially sized local quantum circuits can also not be learned efficiently with respect to a generator. This result can be extended to any universal gate set by virtue of the Solovay-Kitaev theorem.
We obtain the stated depth dependence by use of a rescaling argument trading complexity for depth.
\end{proofsketch}

Previous work has suggested, and provided numerical evidence, that learning a generator for quantum circuit output distributions is hard for classical learning algorithms \cite{coyleBornSupremacyQuantum2020,generativegeneral,rudolph2020generation,niu2020learnability}. Theorem \ref{PRFhard} provides a rigorous proof for this and, interestingly, shows that these distributions are also hard to learn using quantum algorithms -- including QCBM based learners. As such, one cannot hope to use the output distributions of local quantum circuits to prove a probabilistic modelling separation between QCBM  based algorithms and classical algorithms.

We note that our proof technique shares similarities with that of Ref.~\cite{arunachalam2020quantum}, where it was shown that learning Boolean functions generated by constant depth classical circuits is hard for quantum algorithms, even with quantum examples. However classes of Boolean functions which are hard to learn cannot be generically used to create distribution classes which are hard to learn with respect to a generator \cite{xiaoLearning10}. As such, our results do not follow directly from theirs, despite similarities in the proof strategies. 

\smallskip
{\noindent\it \textbf{Hardness of learning with statistical query algorithms.}} --- 
In the previous sections we have established the hardness of learning the output distributions of polynomial depth circuits. However, the efficient learnability of shorter circuits remains open. In this section we show that the hardness results of the previous sections can be strengthened to hold for the output distributions of super-logarithmic depth circuits, if one considers a restricted -- but practically highly relevant -- class of learning algorithms.

To understand this restriction recall that in Theorem \ref{Clifford_learning} we have seen an example of a distribution class -- namely the output distributions of Clifford circuits -- whose intrinsic algebraic structure allowed us to devise an efficient learning algorithm. In particular, this algorithm is able to exploit individual samples from the target distribution, by using the promise that the target distribution is the uniform distribution over some affine subspace of $\mathbb{F}_2^n$. However, in the absence of a strong promise on the structure of the unknown distribution to be learned, it is a-priori unclear how a learning algorithm should utilize individual samples from the target distribution. As such, most \textit{generic} distribution learning algorithms -- i.e., algorithms which are not designed specifically for one particular distribution class -- work by using samples from the unknown distribution to estimate statistical averages with respect to that distribution~\cite{mohamed2017learning}. Indeed, this is the case for almost all gradient based algorithms used in practice, both for classical neural network model classes (such as RBMs and GANs)~\cite{mohamed2017learning} as well as quantum circuit based model classes such as QCBMs~\cite{liuDifferentiableLearningQuantum2018,coyleBornSupremacyQuantum2020}.

In order to formally study the properties of such learning algorithms, we assume that the learning algorithm does not have access to samples from the unknown distribution $P$, but only to approximate statistical averages with respect to $P$. More specifically, we assume that the algorithm has access to a \textit{statistical query oracle}, which when queried with some efficiently computable function $\phi:\{0,1\}^n\rightarrow[-1,1]$ returns some $v$ such that $\left|\,\mathbb{E}_{x\sim P}[\phi(x)]-v\,\right|\leq \tau$ - i.e., an approximation of the expectation value of $\phi$ with respect to $P$, up to accuracy $\tau$~\cite{feldman2017general}. While in principle one could consider any accuracy parameter $\tau$, we consider at most inverse polynomial accuracy -- i.e., $\tau = \Omega(1/\mathrm{poly}(n))$ -- as in this regime the statistical query oracle can be efficiently simulated from samples, and query-complexity lower bounds with respect to statistical queries yield computational complexity lower bounds with respect to sample queries~\cite{diakonikolas2017statistical}.

\begin{restatable}{theorem}{SQhard}
\label{SQhard}
There is no query efficient algorithm for learning from inverse polynomially accurate statistical queries
\begin{itemize}
    \item $\D_\Cl$ at depth $\omega(\log(n))$,
    \item $\D_{\mc G}$ at depth 
    $\omega(\log^k(n))$
    where $k$ is a constant depending on the universal gate set $\mc G$ (which can be as small as $2$), 
\end{itemize}
 with respect to either generators or evaluators.
\end{restatable}

\begin{proofsketch}
As shown in Refs.~\cite{kearns1998efficient, blum_weakly_1994} learning parities in the statistical query model is hard. From this, one can prove that the output distributions of parity functions on uniformly random inputs are also hard to learn from statistical queries. We have already shown in the proof of Theorem \ref{thm:Clifford_T_eval_hardness} that the output distributions of parity functions can be realized by linear depth Clifford circuits. Combining these two facts yields the hardness result for linear depth Clifford circuits. We then obtain the first claim by applying a rescaling argument which trades circuit depth for complexity. We obtain the second claim by using robustness properties of the statistical query oracle, coupled with the Solovay-Kitaev theorem to approximate Clifford circuits.
\end{proofsketch}

A first immediate consequence of the above result is that one cannot hope to use the output distributions of super-logarithmic depth local circuits to prove a practical separation between the power of classical learning algorithms and QCBM's, provided one uses previously-proposed QCBM based learning algorithms based on statistical queries~\cite{liuDifferentiableLearningQuantum2018,coyleBornSupremacyQuantum2020}.  Additionally, Theorem~\ref{PRFhard} leaves open the possibility that there exists some efficient learning algorithm for circuits with depth less than $n^{\Omega(1)}$. However, as hardness in the statistical query model is often taken as evidence for hardness in the sample model~\cite{feldman2017general},  the above result provides evidence that Theorem~\ref{PRFhard} could potentially be strengthened to hold for the output distributions of super-logarithmic depth circuits. At least, any efficient learning algorithm for such circuits must utilize individual samples in a non-trivial way.

\smallskip
{\noindent\it \textbf{Conclusions.}} ---
In this letter, we have provided an extensive characterization of the complexity of learning the output distributions of local quantum circuits. 
Apart from being of fundamental interest in its own right, this characterization also contributes to our understanding of the relationship between the learnability and simulatibility of local quantum circuit output distributions.

Moreover, our results have multiple implications for the emerging field of \textit{quantum machine learning}. In particular, a major focus of current research efforts in this direction is the identification of problems for which one can rigorously prove a separation between the power of quantum and classical learning algorithms \cite{arunachalam2017guest}. Previous work has leveraged cryptographic assumptions to construct highly fine-tuned learning problems for which fault-tolerant quantum computers can obtain an exponential advantage~\cite{liu2021rigorous, Sweke2021quantumversus,jerbi2021variational}. The output distributions of quantum circuits were a primary candidate for establishing a separation for a natural learning problem. However, our work establishes that this is not possible, 
and, therefore, implies the need to identify new strategies for proving practically relevant quantum advantages in machine learning.
In particular, our work complements existing results~\cite{stilckfrancaLimitationsOptimizationAlgorithms2021} that place limitations on the applicability of near-term hybrid quantum-classical learning algorithms, including QCBMs. 

There remain many exciting questions. Firstly, are our worst-case bounds tight? In particular, can one exhibit efficient learning algorithms for the circuit depths not covered by our hardness results? Secondly, can one characterize the \textit{sample} complexity of the learning tasks we have considered. Thirdly, in order to gain insight into the performance of heuristic learning algorithms, it is important to understand the {\em average-case} complexity of learning the output distributions of local quantum circuits. Additionally, it is interesting to study the learnability of other physically-motivated distributions, such as those arising from free-fermionic evolutions~\cite{aaronson2021efficient,aaronson_retract}. Finally, to fully characterize the relationship between simulatability and learnability, it is of interest to understand whether hardness of simulation implies hardness of learning. In particular, are there circuit distributions which are hard to classically simulate, while being efficiently learnable?

\smallskip
{\it Acknowledgments.}
We are thankful for excellent discussions with Matthias Caro, Hakop Pashayan and feedback of an unknown peer reviewer. This work has been funded by the Cluster of Excellence MATH+ (EF1-11), the BMWK (PlanQK), the BMBF (Hybrid, QPIC-1), the DFG (CRC183, EI 519 20-1), the QuantERA (HQCC), the Munich Quantum Valley (K8), and the Alexander von Humboldt Foundation.   

\bibliography{refs}
\bibliographystyle{unsrt}

\onecolumngrid
\newpage

\appendix

\section{Preliminaries}

\noindent We start by giving formal definitions of the objects and problems considered in this work. Throughout we denote by $\mathcal{F}_n$ the set of Boolean functions from $\lrb{0,1}^n$ to $\lrb{0,1}$, by $\D_n$ the set of probability distributions over $\lrb{0,1}^n$. A subset $\D\subset\D_n$ is referred to as a distribution class. For two discrete probability distributions $P,Q:\01^n \rightarrow [0,1]$, we denote by ${\tv(P,Q):= \frac{1}{2} \sum_{x\in \01^n} |P(x)-Q(x)|}$ the total variation distance between them.
The trace distance of two quantum states $\rho$ and $\sigma$ is given by  $\frac{1}{2}\norm{\rho-\sigma}_{\tr}\,$, where $\norm{\,\cdot\,}_{\tr}$ denotes the trace norm.
Access to distributions is formalized by assuming access to some oracle that has a specific operational structure. In particular, we use the sample and the statistical query oracle which are defined as follows.

\begin{definition}[Distribution oracles]\label{d:oracles}
Given $P\in\mathcal{D}_n$, some $\tau \in (0,1)$, we define:
\begin{enumerate}
    \item The sample oracle $\samp(P)$ as the oracle which, when queried, provides a sample $x\sim P$. 
    \item The statistical query oracle $\stat_{\tau}(P)$ as the oracle which, when queried with a function $\phi:\{0,1\}^n\rightarrow [-1,1]$, responds with some $v$ such that $|\Ex_{x\sim P}[\phi(x)] - v| \leq \tau$. 
\end{enumerate}
\end{definition}

\noindent Let us next define generators and evaluators, the central objects of this work, whose learnability we study. Informally, a generator for a given distribution $P$ is an algorithm that generates samples from $P$. Likewise, an evaluator for $P$ is an algorithm that computes $P(x)$ for all $x$ in the support of $P$. More precisely:

\begin{definition}[Generators]\label{d:gen} Given some probability distribution $P\in\mathcal{D}_n$, we say that a probabilistic (or quantum) algorithm $\gen_P$ is a generator for $P$ if $\gen_P$ produces samples according to $P$.
\end{definition}

\begin{definition}[Evaluators]\label{d:eval} Given some probability distribution $P\in\mathcal{D}_n$, we say that
an algorithm $\eval_P:\{0,1\}^n\rightarrow [0,1]$ is an evaluator for $P\in\mathcal{D}_n$ if on input $x\in\{0,1\}^n$ the algorithm outputs $\eval_P(x) = P(x)$. 
\end{definition}

We are interested in learning the output distributions of quantum circuits. To formalize this,
we use the framework for {\em learning a distribution} as introduced in Ref.~\cite{Kearns:1994:LDD:195058.195155}.
This definition is analogous to the definition of probably-approximately correct (PAC) {\em function} learning, in that it introduces parameters $\eps$ and $\delta$ to quantify approximation error and probability of successful approximation, respectively. 

\begin{problem}[$(\eps,\delta)$-distribution-learning]\label{prob:eps-del-learning}
Let $\eps,\delta\in(0,1)$ and let $\D$ be a distribution class. Let $\mathcal{O}$ be a distribution oracle. The following task is called $(\eps,\delta)$-distribution-learning $\D$ from $\mc O$ with respect to a  generator (evaluator): Given access to oracle  $\mc O(P)$ for any unknown $P\in\D$, output with probability at least $1-\delta$ an efficient generator (evaluator) of a distribution $Q$ such that $\tv(P,Q)<\eps$. 
\end{problem}

\begin{definition}[Efficiently learnable distribution classes] Let $\mathcal{D}$ be a distribution class, and let $\mathcal{O}$ be a distribution oracle. We say that $\mathcal{D}$ is computationally (query) efficiently learnable from $\mathcal{O}$ with respect to a generator/evaluator, if there exists an algorithm $\mathcal{A}$ which for all $(\epsilon,\delta) \in (0,1)$ solves the problem of $(\epsilon,\delta)$-distribution learning $\mathcal{D}$ from $\mathcal{O}$ with respect to a generator/evaluator, using $O(\mathrm{poly}(n,1/\epsilon,1/\delta))$ computational steps (oracle queries).
\end{definition}

\noindent As we are most often concerned with computational efficiency and with the sample oracle, we often omit these qualifiers in this case, and simply say ``$\mathcal{D}$ is efficiently learnable". If a distribution class is not efficiently learnable, then we say it is hard to learn.\newline 

We are particularly interested in distribution classes induced by quantum circuit classes by measuring each corresponding quantum circuit in the computational basis. We denote such classes in the following fashion:

\begin{definition}[$\D_{\mc G}(n,d)$]
Let $\mc G$ be a gate set and let $n,d\in\mathbb{N}$. We denote by $\D_{\mc G}(n,d)$ the set of output distributions of $n$-qubit nearest neighbor quantum circuits with gates from the gate set $\mc G$ at depth $d$. In particular, $\D_{\mc G}(n,d)$ contains those distributions $P\in\D_n$ that can be written as
\begin{align}
    P(x)=\abs{\mel{x}{U}{0^n}}^2\,,
\end{align}
where $U$ can be written as a depth $d$ nearest neighbor quantum circuit in one dimension on $n$ qubits composed of gates from $\mc G$.
\end{definition}

\section{Useful reductions}

\noindent In this section we provide a variety of lemmata, used in the proofs of our main theorems. We start with an embedding lemma which, at a high level, allows us to trade circuit depth for computational complexity of learning. More specifically, this lemma allows us to take a lower bound for learning the output distributions of a class of quantum circuits of a given depth, and obtain a new \textit{smaller} lower bound for learning \textit{shorter} quantum circuits. This allows us to take existing lower bounds for some class of circuits, and identify the shortest circuit depth which admits a super-polynomial lower bound. The intuition behind this lemma is illustrated in Fig.~\ref{fig:embedding}, and is as follows: Assume learning the output distributions of a given class of quantum circuits takes at least a certain number of computational steps (or oracle queries). Now consider the class of circuits one obtains by embedding the original circuits into wider circuits, which act trivially on the extra qubits.  Intuitively, learning the output distributions of the wider quantum circuits should take at least the same number of steps (oracle queries) as for the original circuits. However, as a function of the number of qubits, both the depth of the wider quantum circuits, and the computational time (number of oracle queries) required for learning their output distributions, is reduced. We formalize this below:

\begin{lemma}[Embedding reduction]\label{lem:embedding}
Let $n\in\mathbb{N}$, $\eps,\delta\in(0,1)$ and let $\tau(n)>0$ be a function depending on $n$. Let $f,g:\mathbb{N}\rightarrow\mathbb{N}$ be functions where $f$ is monotonous and $g$ is strictly monotonous with $n\leq g(n)$. We call $g$ the stretch. Assume $(\eps,\delta)$-learning $\D_{\mc G}(n, f(n))$ 
\begin{itemize}
    \item with respect to a generator from samples requires at least time  $t(n,\eps,\delta)$, and $g = O(\poly(n))$, or
    \item with respect to any representation requires at least $q(n,\eps,\delta)$ statistical queries with tolerance  $\tau(n)$.
\end{itemize}  
Then it requires at least time $t(g^{-1}(n),\eps,\delta)-O(\poly(n))$ (respectively  $q (g^{-1}(n),\eps,\delta)$ statistical queries with tolerance at least $\tau\circ g^{-1}(n)$) to  $(\eps,\delta)$-learn $\D_{\mc G}(n,f\circ g^{-1}(n))$ with respect to the corresponding representation.
\end{lemma}

\begin{proof}
To begin, we consider the first claim. Let $\A$ be an algorithm that $(\eps,\delta)$-learns $\D_\G(n, f\circ g^{-1}(n))$ from samples with respect to a generator in time $t^*(n,\eps,\delta)$. We now define an algorithm $\B$ that makes use of $\A$ as a subroutine to $(\eps,\delta)$-learn $\D_\G(n,f(n))$ from samples with respect to a generator. As we will show, its runtime is bounded by $t^*(g(n),\eps,\delta)+O(\poly(n))$. 

Let $k\in\mathbb{N}$, denote $n=g(k)$ and let $P\in\D_\G(k,f(k))$ be a distribution to which we are given sample access via $\samp(P)$. We define algorithm $\B$ as follows: $\B$ first emulates a sample oracle $\samp(Q)$ to a distribution $Q\in\D_\G(n,f\circ g^{-1}(n))$ defined as
\begin{align}
    Q(x_1,\dots,x_k,x_{k+1},\dots,x_n)=
    \begin{cases}
        P(x_1,\dots,x_k)\,,\,\text{if}\,x_{k+1}=\dots=x_n=0\\
        0\,,\,\text{else}
    \end{cases}
\end{align}
by appending $n-k$ zeros to any bit string $(x_1,\dots,x_k)$ output by $\samp(P)$. Then $\B$ invokes $\A$ with access to $\samp(Q)$ which returns a generator $\gen_{Q'}$ for a $Q'\in\D_\G(n,f\circ g^{-1}(n))$. $\B$ then returns the generator $\gen_{P'}$ which is defined as follows: Run $\gen_{Q'}$ and receive a sample $(x_1,\dots,x_n)$. Return $(x_1,\dots,x_k)$ discarding the remaining $n-k$ bits.

Let us now analyze the correctness of $\B$: By the tensorial structure of quantum circuits, $\samp(Q)$ is a valid sample oracle to some $Q\in\D_\G(n,f\circ g^{-1}(n)))$. Therefore, $\A$ will with probability at least $1-\delta$ return a generator $\gen(Q')$, efficient in $n$, to some $Q'$ that is at least $1-\eps$ close to $Q$ in $\tv$-distance. Now we observe that $\gen_{P'}$ is a generator for the marginal distribution $P'$ of $Q'$ on the first $k$ bits. Hence, assuming that $Q'$ is a correct $\eps$-approximation to $Q$, by the contractivity of the $\tv$-distance, we find that $P'$ is a valid $\eps$-approximation to $P$. Moreover, since $g(k)=O(\poly(k))$ by assumption, we find that $\gen_{P'}$ is also efficient in $k$. Hence, with probability $1-\delta$ our algorithm $\B$ will find an efficient generator for a distribution that is $\eps$ close to the original distribution $P$, thus proving the correctness.

We now observe that all steps in the reduction can be implemented with an at most polynomial overhead. Hence, learning $\D_\G(k, f(k))$ takes time at most $t^*(n,\eps,\delta)+O(\poly(n))=t^*(g(k),\eps,\delta)+O(\poly(k))$, proving the first claim.\\

\begin{figure}
\includegraphics[scale=.65]{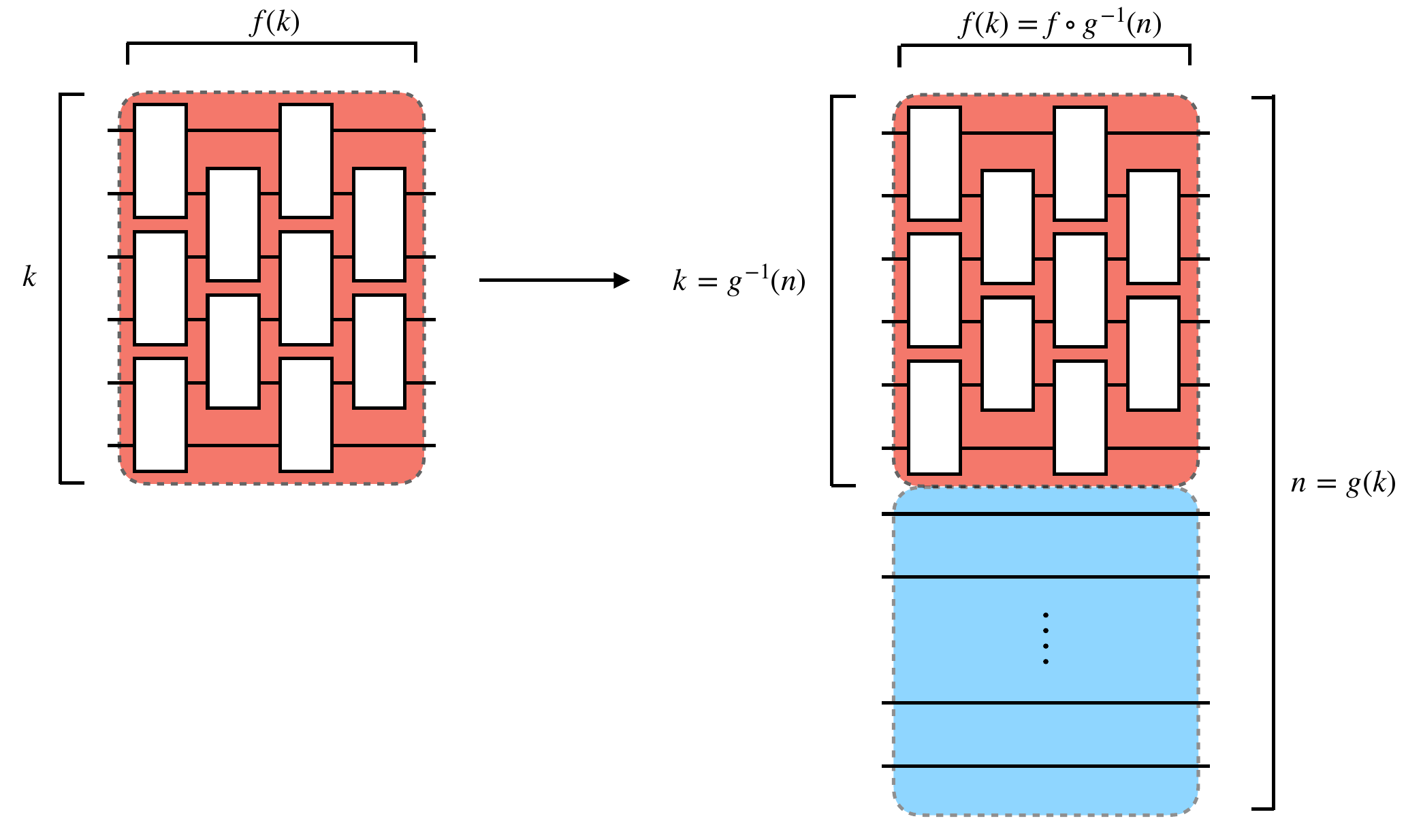}
\caption{Illustration of the embedding reduction used in the proof of Lemma~\ref{lem:embedding}. Given a class of circuits on $k$ qubits, we can define a new class of circuits on $n= g(k)$ qubits by embedding the original circuits onto the first $k$ qubits. Intuitively, the number of computational steps (oracle queries) required to learn the output distributions of the wider circuits, should be at least as many as that required for the original circuits. However, as a function of the number of qubits in the wider circuits, both the depth and the learning complexity are reduced by the inverse of the ``stretch-factor" $g$.}\label{fig:embedding}
\end{figure}

\noindent The second claim follows from a similar reasoning replacing computational time with oracle queries. Since the claim is in terms of the query complexity and as such inherently information theoretic, we do not need to impose the stretch $g$ to be polynomial. Similarly, as the reduction itself does not make any statistical queries we will get the direct mapping of the query complexity $q(n,\eps,\delta)\mapsto q(g^{-1}(n),\eps,\delta)$ when applying $\A$ as a subroutine. Moreover, due to the information theoretic nature of the statement it applies to both generators and evaluators. In particular, it suffices to show the claim for generators, as we can, at least in a computationally inefficient way, obtain the corresponding evaluators without additional statistical queries.

This means, we only need to adapt the oracle emulation: Assume $\phi:\{0,1\}^n\rightarrow[-1,1]$ to be some function queried by $\A$ and let $P$ and $Q$ be as before. To emulate $\stat_{\tau\circ g^{-1}(n)}(Q)$ when queried with $\phi$ we query $\stat_{\tau(k)}(P)$ with $\theta$ and return the corresponding value, where
\begin{align}
    \theta(x_1,\dots,x_k)&=\phi(x_1,\dots,x_k,0,\dots,0)\,.
    \intertext{We complete the proof by noting that}
    \Ex_{x_1,\dots,x_k\sim Q}[\phi(x_1,\dots,x_n)]=\Ex_{x_1,\dots,x_k\sim P}&[\phi(x_1,\dots,x_k,0,\dots,0)]=\Ex_{x_1,\dots,x_k\sim P}[\theta(x_1,\dots,x_k)]\,,
\end{align}
and $\tau\circ g^{-1}(n)=\tau(k)$ such that this prescription is indeed a valid emulation. The correctness proof is identical to that of the first claim.
\end{proof}

The same trade-off of depth for complexity also applies to learning with respect to evaluators. In particular, in the special case of $\eps=0$, we immediately obtain the following corollary.

\begin{corollary}\label{cor:eval-embedding}
Let $n,\delta,g,f$ as before, $g=O(\poly(n))$ and assume that $(0,\delta)$-learning $\D_\G(n, f(n))$ from samples with respect to an evaluator requires at least time $t(n,\delta)$. Then it requires at least time $t(g^{-1}(n),\delta)-O(\poly(n))$ to $(0,\delta)$-learn $\D_\G(n, f\circ g^{-1}(n))$  with respect to an evaluator.
\end{corollary}

\begin{proof}
The proof is identical to the first part of the proof of \Cref{lem:embedding} only that the output of $\A$ is, with probability $1-\delta$, the evaluator of $Q$. Note, as $\eps=0$ it holds that $Q'=Q$ and $P'=P$. In order to transform $\eval_Q$ to the evaluator of the original $P$ we simply map 
\begin{align}
    \eval_P(x_1,\dots,x_k)=\eval_Q(x_1,\dots,x_k,0,\dots,0)\,. \label{eq:eval_reduc}
\end{align}
The correctness follows from the correctness of $\A$ together with $\eps=0$.
\end{proof}

In principle, the proof above also works in the case of non-zero $\eps$. However, the output of the algorithm will in general not be an evaluator in the exact sense of Definition \ref{d:eval}. This is because the mapping in Eq.~\eqref{eq:eval_reduc} does not preserve the normalization of the probability distribution. For practical purposes, however, one can just relax the definition of an evaluator to also apply to non-negative vectors instead of normalized probability distributions and replacing the $\tv$ distance by the $\ell_1$-norm. Then, the above proof goes through for $\eps\neq 0$.

We have stated Lemma \ref{lem:embedding} in its most general form as we believe that it might be of use on its own. In order to give a concrete 
example, we provide a corollary that will also be of use in the proof of Theorem \ref{thm:Clifford_T_eval_hardness}. 

\begin{corollary}\label{ntoomega}
Let $n>0$ and assume $d=O(\poly(n))$. If there is no efficient algorithm for learning $\D_\G(n, d(n))$ with respect to a generator, then there is no efficient algorithm for learning $\D_\G(n, d'(n))$ with respect to a generator for any $d'=n^{\Omega(1)}$.
\end{corollary}

\begin{proof}
Let $r\in\mb N$ be such that $d(n)=O(n^r)$. Then, via
Lemma \ref{lem:embedding} with $g(n)=n^{rs}$ for some $s\in\mb N$, we find that there is no efficient algorithm for learning $\D_\G(n, d'(n))$ with $d'(n)=\Omega(n^{1/s})$. 
The claim then follows since $s\in\mb N$ is arbitrary.
\end{proof}

Next we will clarify in which way hardness results for one learning problem can be leveraged to obtain hardness results for a different distribution class which only approximates the former. 
This is a crucial tool for lifting lower bounds for circuits with some specific gate set, to generic universal quantum circuits, since the latter are known to efficiently approximate the former due to the Solovay-Kitaev theorem. Let us start by introducing some notation.

\begin{definition}
Let $\D$ and $\mc H$ be distribution classes over the domain $X$ and let $\sigma\in[0,1)$. We say $\D$ is $\sigma$-approximately contained in $\mc H$ (with respect to the total variation distance), and write
\begin{align}
    \D\subseteq_\sigma\mc H\,,
\end{align}
if for every $P\in\D$ it exists a $Q\in\mc H$ such that $\tv(P,Q)\leq\sigma$.
\end{definition}
Given this we find the following reduction from the learnability of a class $\mc H$ to the learnability of a approximately contained class $\D$. Alternatively, this implies that a harndess result on $\D$ translates to a corresponding hardness result on $\mc H$.

\begin{lemma}[Approximation reduction]\label{lem:approximation-reduction}
Let $\D$, $\mc H$ and $\sigma$ as before and $\D\subseteq_\sigma\mc H$. Assume that $\mc H$ is $(\eps,\delta)$-learnable from $s$ samples with respect to a representation. Then $\D$ is $(\eps+\sigma,\delta+s\sigma)$-learnable from $s$ samples with respect to the same representation.
\end{lemma}

\begin{proof}
Let $\A$ be an algorithm that $(\eps,\delta)$-learns $\mc H$ with respect to a representation from $s$ samples. Then, applying $\A$ to $\D$ directly yields an $(\eps+\sigma,\delta+s\sigma)$-learner with respect to the same representation. We first show this assuming $\A$ to be deterministic before switching to the general case. Assume $\A$ to be deterministic. For
any $P\in\mc H$ define the event
\begin{align}\label{eq:event-definition}
    \mc E(P,\eps,\A):=\{(x_1,\dots,x_s)\mid\tv\lr{\A(x_1,\dots,x_s),P}<\eps\}\subseteq X^s\,.
\end{align}
We interpret the characteristic function $\mc E(x_1,\dots,x_s):=\mathds{1}_{\mc E(P,\eps,\A)}(x_1,\dots,x_s)$ as a random variable with respect to a distribution over $X^s$.

Since $\D\subseteq_\sigma\mc H$ we know that for every $Q\in\D$ there exists a $P\in\mc H$ such that $\tv(P,Q)\leq\sigma$. This implies
\begin{align}\label{eq:tensortvdistance}
    \tv(P^{\otimes s},Q^{\otimes s})\leq s\sigma\,.
\end{align}
Therefore, it must hold
\begin{align}\label{eq:approximateestimation}
    &\Pr_\A\lrq{\tv\lr{\A^{Q},P}<\eps}
    =\Pr_{(x_1,\dots,x_s)\sim Q^{\otimes s}}\lrq{\mc E}
    \geq\Pr_{(x_1,\dots,x_s)\sim P^{\otimes s}}\lrq{\mc E}
    -s\sigma
    =\Pr_\A\lrq{\tv\lr{\A^{P},P}<\eps}-s\sigma
    =1-\delta-s\sigma\,,
\end{align}
where $\A^Q$ ($\A^P$) is short hand notation for the output of the algorithm $\A$ with oracle access to $\samp(Q)$ ($\samp(P)$). The inequality is due to Eq.~(\ref{eq:tensortvdistance}) and the variational characterization of the $\tv$-distance. Hence, running $\A$ on any $Q\in\D$ will, with probability $1-(\delta+s\sigma)$ return a representation of some $P'$ with
\begin{align}
    \tv\lr{P',Q}\leq\tv\lr{P', P}+\tv\lr{P,Q}\leq\eps+\sigma\,,
\end{align}
proving the deterministic case.\\

Now assume $\A$ to be a random algorithm. Thus, the randomness in the second part of Eq.~(\ref{eq:approximateestimation}), $(x_1,\dots,x_s)\sim Q^{\otimes s}$, gets replaced by $(x_1,\dots,x_s,r_1,\dots,r_k)\sim Q^{\otimes s}\otimes D$ (and similarly for $P$) where $D$ is the distribution on $k$ bits induced by derandomizing $\A$. The claim then follows from the same argument replacing 
Eq.~(\ref{eq:tensortvdistance}) by
\begin{align}
    \tv(Q^{\otimes s}\otimes D,P^{\otimes s}\otimes D)\leq s\sigma\,,
\end{align}
which follows from Eq.~(\ref{eq:tensortvdistance}) and the factorization of the $\ell_1$-norm.

The argument is the same in spirit for quantum algorithms, only the formulation of the `derandomization' procedure changes due to the quantum nature of the algorithm. Recall, that any quantum algorithm $\A$ which makes $s$ queries to $\samp(Q)$ can be written as a quantum circuit acting on a suitable input density matrix encoding the $s$ queries to the oracle and the internal quantum resources of the algorithm in terms of an auxiliary density matrix. We, 
therefore, replace $Q^{\otimes s}\otimes D$ from the previous reasoning by $\rho_Q^{\otimes s}\otimes\rho_\A$ (and likewise for $P$) where $\rho_Q=\sum_{i\in X}Q(i)\ketbra i$ is the diagonal mixed state corresponding to $Q$ and $\rho_\A$ is the density matrix corresponding to $\A$'s auxiliary space. 
Using the factorization of the trace norm and the fact that both $\rho_Q$ and $\rho_P$ are diagonal in the computational basis we find
\begin{align}
    \frac12\norm{\rho_Q^{\otimes s}\otimes\rho_\A-\rho_P^{\otimes s}\otimes\rho_\A}_{\tr}=
    \frac12\norm{\rho_Q^{\otimes s}-\rho_P^{\otimes s}}_{\tr}=
    \tv(P^{\otimes s}, Q^{\otimes s})\leq s\sigma\,.
\end{align}
Hence, the claim follows from 
\begin{align}
    \Pr_{\A}[\tv(\A^{Q},P)<\eps]
    =\Tr[\Pi\cdot\rho_Q^{\otimes s}\otimes\rho_\A]
    \leq\Tr[\Pi\cdot\rho_P^{\otimes s}\otimes\rho_\A]-s\sigma
    =\Pr_{\A}[\tv(\A^{P},P)<\eps]-s\sigma\,,
\end{align}
where $\Pi$ is the POVM encoding the application of the quantum algorithm $\A$ to the input $\rho^{\otimes s}\otimes\rho_\A$ and then projecting onto the valid solutions.
\end{proof}

For the proof of Theorem \ref{SQhard} we will need a slightly adjusted version of Lemma \ref{lem:approximation-reduction} which applies to the setting of statistical query learning.

\begin{lemma}[Statistical query approximation reduction]\label{lem:stat-query-approximation-reduction}
Let $\D,\mc H$ be distribution classes over the same domain $X$ , let $\tau>\sigma>0$ and let $0<\gamma<\tau-\sigma$. Assume $\D\subseteq_\sigma\mc H$ and that $\mc H$ is $(\eps,\delta)$-learnable from $q$ statistical queries with tolerance $\tau$. Then $\D$ is $(\eps+\sigma,\delta)$-learnable from at most $q$ statistical queries with tolerance $\gamma$.
\end{lemma}

\begin{proof}
The proof idea is similar to that of Lemma \ref{lem:approximation-reduction} though, by properties of statistical query learning, is technically much simpler.
To begin with assume $\A$ to be an algorithm that $(\eps,\delta)$-learns $\mc H$ with $q$ statistical queries of tolerance $\tau$. Then, applying $\A$ to $\D$ directly yields an $(\eps+\sigma,\delta)$-learner for $\D$ which uses at most $q$ statistical queries of tolerance $\gamma\leq\tau-\sigma$.
To see this, we first note that by assumption, for any $P\in\D$ there exists a $Q\in\mc H$ such that $\tv(P,Q)<\sigma$. 
By the variational characterization of the total variation distance and the triangle inequlity we hence find that, for any $v\in[-\gamma,\gamma]$ 
\begin{align}
    \abs{\Ex_{x\sim P}[\phi(x)]+v-\Ex_{x\sim Q}[\phi(x)]}\leq \abs{\Ex_{x\sim P}[\phi(x)]-\Ex_{x\sim Q}[\phi(x)]} + \abs{v} <\sigma+\tau-\sigma=\tau\,.
\end{align}
Thus, any oracle $\stat_\gamma(P)$ can be interpreted as a $\stat_\tau(Q)$ oracle. This implies, that when run with access to $\stat_\gamma(P)$ algorithm $\A$ will, with probability at least $1-\delta$ return a representation for some distribution $D$ that is $\eps$ close to $Q$ in total variation distance. By the triangle inequality this is at most $\eps+\sigma$ far from $P$ completing the proof. 
\end{proof}

\noindent As we are exclusively concerned here with distribution classes associated with local quantum circuits, the following additional standard results will be useful to us to quantify the extent to which the output distributions of one class of quantum circuits can be approximated by the output distributions of another class of quantum circuits. 

\begin{lemma}\label{lem:tracenorm}
Let $\rho=\ketbra\psi$ and $\sigma=\ketbra\phi$ be pure quantum states. Then it holds
\begin{align}
    \norm{\rho-\sigma}_{\tr}=2\sqrt{1-\abs{\braket{\psi}{\phi}}^2}\,.
\end{align}
\end{lemma}

\begin{proof}
(From the proof of Theorem 10 in Ref.~\cite{brandao_models_2021}.) Denote $X=\rho-\sigma$. Then $X$ is self-adjoint and $\tr [X]=0$. Hence $X$ has eigenvalues $\lambda$ and $-\lambda$. Moreover $\tr[X^2]=2\lambda^2=2(1-\abs{\braket{\psi}{\phi}}^2)$. Hence, $\lambda=\sqrt{1-\abs{\braket{\psi}{\phi}}^2}$ and the claim follows from $\norm{X}_{\tr}=2\abs{\lambda}$.
\end{proof}

\begin{lemma}\label{lem:tvapproximation}
Let $n\in\mathbb{N}$. Let $U$ and $W$ be unitary circuits on $n$ qubits and let $P$ and $Q$ be the Born distributions corresponding to $U\ket{0^n}$ and respectively $W\ket{0^n}$. Assume $\norm{U-W}_{\text{op}}<\eps$. Then it holds $\tv(P,Q)<\eps$.
\end{lemma}

\begin{proof}
Denote by $\rho=U\ketbra{0^n}U^\dagger$,  $\sigma=Q\ketbra{0^n}Q^\dagger$ and for any $M\subseteq\{0,1\}^n$ let $\Pi_M=\sum_{i\in M}\ketbra{i}$. Then by the variational characterization of the trace- and total variation distances it holds
\begin{align}
    \tv(P,Q)=\sup_{M}\abs{P(M)-Q(M)}=\sup_{M}\abs{\tr[\rho\Pi_M]-\tr[\sigma\Pi_M]}
    \leq
    \frac{1}{2}\norm{\rho-\sigma}_{\tr}\,.
\end{align}
To estimate the last expression we write
\begin{align}
    \norm{U\ket{0^n}-Q\ket{0^n}}_2=\sqrt{2-2\mathrm{Re}\lr{\mel{0^n}{Q^\dagger U}{0^n}}}<\eps\,,
\end{align}
such that
\begin{align}
    \sqrt{2-2\abs{\mel{0^n}{Q^\dagger U}{0^n}}}\leq\sqrt{2-2\mathrm{Re}\lr{\mel{0^n}{Q^\dagger U}{0^n}}}<\eps\,.
\end{align}
We can now combine this 
with Lemma \ref{lem:tracenorm} to obtain
\begin{align}
    \norm{\rho-\sigma}_{\tr}=2\sqrt{1-\abs{\mel{0^n}{Q^\dagger U}{0^n}}^2}
    <
    2\sqrt{\eps^2-\eps^4/4}\leq2\eps\,,
\end{align}
and hence $\tv(P,Q)<\eps$
\end{proof}

\begin{corollary}[Solovay-Kitaev reduction]\label{cor:solovay-kitaev}
Let $n,d\in\mathbb{N}$, let $\eps>0$ and let $\mc G$ be a universal gate set. Then there exists a constant $c$ such that 
\begin{align}
    \D_{\mathrm{U}(4)}(n,d)\subseteq_\eps\D_{\mc G}(n, d')\quad\text{with}\quad d'=d\cdot\log^c\lr{\frac{n\cdot d}{\eps}}\,.
\end{align}
\end{corollary}

\begin{proof}
By the Solovay-Kitaev theorem \cite{dawson_solovay-kitaev_2005} there exists for any depth $d$ circuit $U$ with at most $n\cdot d$ gates a depth $d'$ circuit $Q$ consisting of gates from $\mc G$ that approximates $U$ in operator norm $\norm{U-Q}_{\text{op}}<\eps$. Hence, applying Lemma
\ref{lem:tvapproximation} yields the claim.
\end{proof}

\noindent Note that there exist universal gate sets for which $c=1$ in the statement of
Corollary \ref{cor:solovay-kitaev} \cite{harrow_efficient_2002}.

\section{Proof of \Cref{Clifford_learning}}
 
\noindent As the proof of \Cref{Clifford_learning} will be based on the algebraic structure of Clifford circuits let us review the following properties first.

\begin{definition}[Affine subspace]\label{def:affine}
An affine subspace $A \subseteq \mathbb{F}_2^n$ is a set such that for every $a,b,c\in A$ and $\lambda\in\mb F_2$ it holds 
\begin{align}
    a+(b-a)+\lambda\cdot(c-a)\in A\,,
\end{align}
where all operations are with respect to $\mb F_2^n$. 
\end{definition}

In other words, for every $a\in A$ the set $A-a$ forms a linear subspace $L$ and $A$ is the set resulting from shifting $L$ by $a$.
This is, there exists an integer $m\leq n$ such that for any $t\in A$ there exists a full-rank matrix $\mbf{R} \in \mb F_2^{m\times n}$, such that
\begin{equation}\label{eq:cliff}
    A = \{\mathbf{R}b + t\,|\, b\in\mathbb{F}^m_2\}\,.
\end{equation}
We say $A$ has dimension $m$. The choice of $\mbf R$ is not unique.

 The output states of Clifford circuits are called stabilizer states.  
 As shown in Refs.~\cite{Dehaene_2003,montanaro2017learning}, up to a global phase, all $n$-qubit stabilizer state vectors $\ket\psi$ can be written in the computational basis as
\begin{equation}
    \ket\psi = \frac{1}{\sqrt{\abs{A}}}\sum_{x\in A} (-i)^{l(x)}(-1)^{q(x)}|x\rangle,
\end{equation}
where $A$ is some affine subspace of $\mathbb{F}^n_2$ and $l,q$ are linear and quadratic functions on $\mathbb{F}^n_2$, respectively.
Thus, we find the following corollary.

\begin{corollary}\label{cor:stabilizer-affine}
For any $P\in\D_\Cl$ there exists an affine subspace $A\subseteq\mb F_2^n$ such that $P=U_A$, where $U_A$ is the uniform distribution on $A$
\begin{equation}
    U_A(x)=
    \begin{cases}
        2^{-d}\,,&\quad d=\dim(A),\; x\in A\\
        0\,,&\quad\text{else.}
    \end{cases}
\end{equation}
\end{corollary}

\noindent For the proof of \Cref{Clifford_learning}, we can make use of the following fact (c.f. Ref.~\cite{Ferreira2013Rank}).

\begin{lemma}\label{lem:linear-subspace-samples}
Let $L\subseteq\mb F_2^n$ be a $m$-dimensional linear subspace with $m\leq n$. Let $x_1,\dots,x_k$ be $k\geq m$ vectors sampled uniformly at random from $L$. Then it holds
\begin{align}
    \Pr\lrq{\mathrm{span}\{x_1,\dots,x_k\}=L}\geq 1-2^{m-k}\,.
\end{align}
\end{lemma}

\noindent \Cref{lem:linear-subspace-samples} can be exploited to learn the affine subspace $A$ from $U_A$ as explained in \Cref{alg:aff_rec2}. This algorithm is a variant of the more general \textit{closure algorithm}, which has previously been used to efficiently solve on-line learning problems such as learning parity functions and integer lattices 
\cite{helmbold1992learning,auer1994online} and which is used as a subroutine for subexponentially learning parities with noise \cite{Blum2003NoisetolerantLT}. The guarantees of Algorithm \ref{alg:aff_rec2} are as follows.

\begin{algorithm}[H]
  \caption{Affine subspace recovery from samples.
    \label{alg:aff_rec2}}
  \begin{algorithmic}[1]
  	\Statex Input: $\delta\in(0,1)$ and access to $\samp(U_A)$ for some affine subspace $A\subseteq{\mathbb{F}^n_2}$,
    \Statex
    \State Let $k :=n + \lceil\log(1/\delta)\rceil$. 
    Obtain samples $\{x_1,\ldots,x_k\}\sim U_A$ by querying $\samp(U_A)$.
    \State Transform the samples $x_1,\ldots,x_k$ to $y_1,\ldots,y_k$ via $y_i = x_i+ x_1$. \label{al:transform}
    \State Use Gaussian elimination to determine from $y_1,\ldots,y_k$ a maximal linearly independent subset of vectors $V :=\{y_{i_1},\ldots,y_{i_m}\}$.\label{al:lin_find}
    \State Form the full rank $n\times m$ matrix $\mathbf{R}$ by placing vectors from $V$ as columns.
    \State Output $(\mathbf{R},x_1)$.
  \end{algorithmic}
\end{algorithm}

\begin{lemma}[Efficient recovery of affine subspaces] \label{l:affine_sub_recovery2}%
Let $A \subseteq \mathbb{F}^n_2$, $\delta\in(0,1)$ be as stated above. Algorithm \ref{alg:aff_rec2} runs in time $O(\mathrm{poly}(n,1/\delta))$ and uses $O(\mathrm{poly}(n,1/\delta))$ samples, and outputs, with probability at least $1-\delta$, a tuple $(\mathbf{R},t)$ which parametrizes $A$.
\end{lemma}

\begin{proof} The sample complexity is as stated in Algorithm \ref{alg:aff_rec2}. The time complexity follows from the fact that Gaussian elimination on an $n\times m$ matrix, $m<n$, takes time polynomial in $n$. It remains to prove the correctness of this algorithm.

Let $\mbf R'$ be such that $(\mbf R', x_1)$ parametrizes $A$.
Line 2 transforms each $x_i \in A$ into a vector $y_i \in L$ where $L :=\{\mbf{R}'b\,|\,b\in\mathbb{F}^m_2\}$ is the linear subspace in $A$ shifted by $x_1$.
By assumption the original samples $\{x_1,\dots x_k\}$ are uniform on $A$. 
A linear transformation of a uniform distribution is another uniform distribution, such that the new samples $\{y_1,\dots , y_k\}$ are uniform on $L$.
From 
\Cref{lem:linear-subspace-samples} we obtain
\begin{align}
    \mathrm{Pr}\lrq{\mathrm{span}\{y_1,\dots,y_k\}=L} &\geq 1 - 2^{m-k} 
    \geq 1 - 2^{n-k}
    \geq 1 - 2^{n - (n+\log(1/\delta))} 
    =1 - \delta.
\end{align}
Hence, with probability at least $1-\delta$, the columns of $\mathbf{R}$ defined in Step 4 provide a basis for $L$. 

To finish the proof assume that $\mbf R$ is full rank and denote by $A'$ the affine subspace parametrized by $(\mbf R, x_1)$. Then for every $b\in\mb F_2^m$ it holds
\begin{align}
    \mbf R\cdot b+x_1\in \mathrm{span}\{x_1,\dots, x_k\}\subseteq A\,
\end{align}
and hence $A'\subseteq A$. Contrarily, since $\mbf R$ has full rank 
$\abs{A}=\abs{A'}$. Thus $A=A'$, which completes the proof.
\end{proof}

\noindent We now combine these insights to prove the actual statement.

\CliffordLearning*

\begin{proof}
By \Cref{cor:stabilizer-affine}, all distributions in $\mathcal{D}_{Cl}$ take the form of $U_A$  for some affine subspace $A\subseteq\mathbb{F}^n_2$. 
Using \Cref{alg:aff_rec2} in conjunction with \Cref{l:affine_sub_recovery2} we obtain, with probability $1-\delta$, a parametrization $(\mathbf{R},t)$ of $A$ in time $\mathrm{poly}(n,1/\delta)$ using $\mathrm{poly}(n, 1/\delta)$ many samples from $U_A$.

We now get an efficient generator for $U_A$ by uniformly at random sampling $b\sim\mb F_2^m$ and outputting $\mbf R\cdot b+t$. An efficient evaluator that computes $U_A(x)$ on input $x$ is defined as follows: use Gaussian elimination in order to  decide whether $x-t\in\mbf R\mb F_2^m$. If this is the case return $2^-d$ with $d=\dim(A)$. Else return $0$.
Thus it is sample- and computationally-efficient to $(\eps,\delta)$-learn $\D_\Cl$ with respect to a generator and evaluator. 
\end{proof}

\section{Proof of \Cref{thm:Clifford_T_eval_hardness}}

\CliffordPlusT*

\begin{proof}
For each string $s\in\{0,1\}^k$ let $\chi_{(s,k)} \in\mathcal{F}_k$ be the associated parity function on $k$ bits -- i.e. $\chi_{(s,k)}(x) = x\cdot s$ for all $x\in \{0,1\}^k$. For any $\eta\in (0,1/2)$ we define the ``noisy parity distribution on $k+1$ bits" $P_{(s,\eta,k)}\in\mathcal{D}_{k+1}$ via 
\begin{equation}
    P_{(s,\eta,k)}(x,y) = \begin{cases}
        2^{-k}\cdot(1-\eta)\,,& \text{if}\quad y=\chi_{s,k}(x)\\
        2^{-k}\cdot\eta \,,&  \text{else}\,,
    \end{cases}
\end{equation}
for all $s,x\in\{0,1\}^k$. 
Define the distribution $T_l$ as the trivial distribution on $l$ bits - i.e. the distribution with $T_l(0^{l})=1$
and
for any $k\leq n$ define $\mathcal{D}_\eta(n,k)\subseteq \mathcal{D}_{n+1}$ as the set of ``noisy parity distributions on the first $k+1$ bits" -- i.e. $\mathcal{D}_\eta(n,k) = \{P_{(s,\eta,k)}\otimes T_{n-k}\,|\, s\in \{0,1\}^k\}$. 

In the proof of Theorem 16 in Ref.~\cite{Kearns:1994:LDD:195058.195155}, the authors show that, under the LPN assumption, there is no efficient algorithm for learning the noisy parity distributions $\mathcal{D}_\eta(n,n)$ with respect to an evaluator, for any $\eta\in (c,1/2-c)$ where $c\in \Omega(1)$. In other words, in this parameter range, all algorithms for learning $\mathcal{D}_\eta(n,n)$ with respect to an evaluator require
$\omega(\mathrm{poly}(n))$ time.  By using similar reasoning to that used in the proof of Lemma~\ref{lem:embedding} -- i.e. embedding the noisy parity distributions onto a subset of bits -- one can extend this result to show that, assuming the LPN assumption, any algorithm for learning $\mathcal{D}_\eta(n,k)$ with respect to an evaluator requires $\omega(\mathrm{poly}(k))$ time
. As $\omega(\mathrm{poly}(n^{\Omega(1)})) = \omega(\mathrm{poly}(n))$ we can conclude that, assuming the LPN assumption, there exists no efficient 
algorithm for learning $\mathcal{D}_\eta(n,n^{\Omega(1)})$ with respect to an evaluator.

Next, we note that for any $s\in \{0,1\}^k$, when $\eta=\sin^2(\pi/8)\approx 0.146$, the distribution $P_{(s,\eta,k)}$ is the output distribution of the quantum circuit on $k+1$ qubits given in Fig.~\ref{fig:parity_with_noise} with the $\mathrm{CNOT}$ gates between the $i^{th}$ and the $(k+1)^{st}$ qubit for all $s_i=1$.
As such, for any $k\leq n$ and any $s\in \{0,1\}^k$, when $\eta=\sin^2(\pi/8)$ the distribution $P_{(s,\eta,k)}\otimes T_{n-k}$ is the output distribution of the quantum circuit on $n+1$ qubits, with the above mentioned circuit from Fig.~\ref{fig:parity_with_noise} on the first $k+1$ qubits, and no gates on the remaining $n-k$ wires. While this circuit contains non-local two-qubit gates, we note that \textit{any} Clifford unitary $U\in\mathrm{Cl}(2^k)$ can be implemented exactly using a depth $d=O(k)$ nearest-neighbour Clifford circuit \cite{Bravyi_2021}. By recompiling the circuit on the first $k+1$ qubits in this way, we obtain an $O(k)$ depth local ``Clifford + one $T$" circuit whose output distribution is $P_{(s,\eta,k)}\otimes T_{n-k}$. Using this, the theorem statement follows from the previously established hardness of learning $\mathcal{D}_\eta(n,n^{\Omega(1)})$ with respect to an evaluator, when $\eta=\sin^2(\pi/8)$. 
\end{proof}

\section{Proof of \Cref{PRFhard}}

\noindent We start by defining the notion of a \textit{pseudorandom} function whose existence is the primary assumption used for Theorem~\ref{PRFhard}. For more detailed definitions and discussion of these objects, see Refs.~\cite{zhandry2012construct,Sweke2021quantumversus}.

\begin{definition}[Classical-secure and standard-secure pseudorandom functions] Let $C\subseteq \mathcal{F}_n$ be a set of efficiently computable functions. We say that $\mathcal{C}$ is a classical-secure (standard-secure) pseudorandom function if for all classical-probabilistic (quantum) polynomial time algorithms $\mathcal{A}$, all polynomials $p$, and all sufficiently large $n$, it holds that
\begin{equation}
    \left|\,\mathrm{Pr}_{f\sim \mathcal{C}}\left[\mathcal{A}^{\mq(f)} = 1 \right] - \mathrm{Pr}_{g\sim \mathcal{F}_n}\left[\mathcal{A}^{\mq(g)} = 1 \right] \,\right| < \frac{1}{p(n)},
\end{equation}
where $\mq(f)$ denotes the membership query oracle, which, when queried with some $x\in\{0,1\}^n$ returns $f(x)$.
\end{definition}

\noindent At a high level, the above definition says that a set of functions $\mathcal{C}$ is classical-secure (standard-secure) if no classical (quantum) algorithm can, with non-negligible probability, distinguish functions drawn uniformly from $\mathcal{C}$ from functions drawn uniformly from $\mathcal{F}_n$. We note that the assumed existence of both classical-secure and standard-secure pseudorandom functions is standard in cryptography~\cite{bogdanov2017pseudorandom, goldreich2009foundations}. We can now recollect the statement of Theorem~\ref{PRFhard}.

\PRFhard*

\begin{proof} Let $\mathcal{C} \subseteq \mathcal{F}_n$ be a classical-secure (standard-secure) pseudorandom-function. Define $\mathcal{D} = \{P_f\,|\, f\in \mathcal{C}\}$ where
\begin{equation}
    P_{f}(x,y) = 
    \begin{cases}
        2^{-n} \,,&\,\text{if}\quad y=f(x) \\
        0\,,&\,\text{else.}
    \end{cases}
\end{equation}
In Ref. \cite{Kearns:1994:LDD:195058.195155} Theorem 17 the authors show that $\mathcal{D}$ is not efficiently classically learnable with respect to a generator, assuming that $\mc C$ is classical-secure. Their hardness result can be straightforwardly extended to apply to quantum learning algorithms as well by requiring $\mathcal{C}$ to be standard-secure.
To leverage their result to show hardness for the output distributions of quantum circuits, we will show how to embed $\mathcal{D}$ into a suitable class of quantum circuits. 
To do so recall that any classical Boolean circuit can be implemented as a quantum circuit via the standard implementation of reversible classical gates together with uncomputation (see Chapter 3 of
Ref.~\cite{nielsenchuang}). For a polynomial size classical circuit, this might incur at most a polynomial overhead in the number of ancilla qubits necessary. Hence, for all $f \in\mc C\subseteq\mc F_n$ there exists a polynomial size quantum circuit $C_f$ on $\poly(n)$ many qubits whose
output distribution is $\tilde P_f$ with
\begin{equation}
    \tilde{P}_{f}(x,y,z) =
    \begin{cases}
        2^{-n}\,,\,&\text{if}\quad z=f(x) \text{ and } y=0^m\\
        0\,,\,&\text{else}\,,
    \end{cases}
\end{equation}
where $m=O(\poly(n))$. Note that any such quantum circuit $C_f$ can be turned into a nearest-neighbor circuit by qubit routing techniques such as using SWAP gates. This will again incur only a polynomial overhead in both size and depth of the circuit. Denote by $\tilde{\mc D}=\{\tilde P_f\mid f\in\mc C\}\subset\mc D_{n+m+1}$ the class of all such distributions. 
Since $n+m+1=O(\poly(n))$ we find that $\tilde{\mc D}$ is hard to learn. 

Lastly, note that $\tilde{\mc D}$ is a subset of the set of the output distributions of polynomial depth quantum circuits.
As such, the output distributions of polynomial depth quantum circuits are not efficiently learnable with respect to a generator. We can see that this holds irrespective of the gate set used (as long as it is universal) by combining Corollary \ref{cor:solovay-kitaev} and Lemma \ref{lem:approximation-reduction}. Finally, using Corollary \ref{ntoomega}, we see that already for $n^{\Omega(1)}$ deep circuits there cannot exist any efficient classical (quantum) algorithm for learning the output distribution with respect to a generator.
\end{proof}

\section{Proof of \Cref{SQhard}}

\noindent Before proving \Cref{SQhard}, we recall a connection of distribution and Boolean function statistical query oracles.

\begin{definition}[Boolean function statistical query oracle \cite{kearns1998efficient}]
Let $f\in\mc F_n$ be a Boolean function, $\tau\in(0,1)$ and let $P\in\D_n$ be a distribution. The Boolean function statistical query oracle of $f$ with respect to $P$ and tolerance $\tau$ is defined as the oracle $\stat_{\tau,P}(f)$ that, when queried with a function $\phi:\{0,1\}^{n+1}\rightarrow[-1,1]$ returns some $v$ such that $\abs{\Ex_{x\sim P}[\phi(x,f(x))]-v}\leq \tau$.
\end{definition}

\begin{corollary}
Let $f\in\mc F_n$ be a Boolean function and let $P\in\D_n$ be a distribution. Define the distribution $P_f\in\D_{n+1}$ as 
\begin{equation}\label{parity_dist}
    P_f(x,y) = 
    \begin{cases}
        P(x)\,,\,&\text{if}\quad y=f(x)\\
        0\,,\,&\text{else}\,.
    \end{cases}
\end{equation}
Then, for any $\tau\in(0,1)$ any statistical query oracle $\stat_\tau(P_f)$ is a Boolean function statistical query oracle $\stat_{\tau,P}(f)$ and vice versa. 
\end{corollary}

\noindent Now we are able to prove the theorem.

\SQhard*

\begin{proof}
To prove the first claim we will reduce statistical query learning of parity functions to statistical query distribution learning of Clifford distributions.
For each string $s\in\{0,1\}^n$ let $\chi_s \in\mathcal{F}_n$ be the associated parity function, and let $\mathcal{C} = \{\chi_s\,|\,s\in\{0,1\}^n\}$ be the class of parity functions. 
For each $s\in\{0,1\}^n$ define $P_s=P_{\chi_s}$ and denote by $\D=\{P_s\mid s\in\{0,1\}^n\}\subset\D_{n+1}$ the class of parity distributions.

As shown in the seminal work of 
Refs.~\cite{kearns1998efficient, blum_weakly_1994}, any algorithm with Boolean function statistical query access of tolerance $\Omega(2^{-n/3})$ to $\mc C$, requires at least $\Omega(2^{n/3-1})$ queries for learning the class of parity functions with respect to the uniform distribution, for any failure probability less than $1/2 - O(2^{-3n})$. 

We now show that, for any $\epsilon < 1/2$, a statistical query algorithm for $(\epsilon,\delta)$-learning $\mathcal{D}$ with respect to an evaluator or generator from $q$ many queries implies a statistical query algorithm for $(0,\delta)$-PAC learning $\mathcal{C}$ from $q$ many queries. 
Assume there exists an algorithm $\mathcal{A}$ for $(\epsilon,\delta)$-learning $\mathcal{D}$ with respect to an evaluator or generator from $q$ many queries to $\stat_\tau(P)$ and with $\eps<1/2$. 
We then define the algorithm $\A'$ which when given access to $\stat_{\tau,U}(\chi_s)$, for some unknown $s\in\{0,1\}^n$, does the following: 
\begin{enumerate}
    \item $\A'$ runs learning algorithm $\A$ where any query to $\stat_\tau(P_s)$ is simulated by querying $\stat_{\tau,U}(\chi_s)$. 
    After at most $q$ queries, $\A$ will output an evaluator $\eval_Q$ (a generator $\gen_Q$) for some distribution $Q\in\D_{n+1}$, which, with probability at least $1-\delta$, is at most $\epsilon$ far from $P_s$.
    \item $\A'$ uses $\eval_Q$ ($\gen_Q$) to find $s$ by brute force. This can be achieved by iterating through all strings $s\in\{0,1\}^n$, test $\tv(P_s,Q)<1/2-\eps$ and return $s$ if true. While this step is not computationally efficient, it requires no additional queries to the oracle.
\end{enumerate}
To see that $\A'$ is correct we note that any two parity distributions are  $\tv(P_s,P_t)=1/2$ far apart. Now, assume $\A'$ runs with access to $\chi_s$, then we find with probability $1-\delta$ that $\tv(Q,P_s)<\eps$. Therefore, any $t$ with $\tv(P_t,Q)<1/2-\eps$ also fulfills $\tv(P_s,P_t)\leq\tv(P_s,Q)+\tv(Q,P_t)<1/2$, where we used $\eps<1/2$. This implies $t=s$. Since all queries in $\A'$ are due to $\A$ we conclude that $\A'$ is an $(0,\delta)$ statistical query learner for $\mathcal{C}$ with respect to the uniform distribution, which requires at most $q$ queries.
We conclude that for any $\epsilon < 1/2$ and $\delta< 1/2$, the problem of $(\epsilon,\delta)$-distribution learning $\mathcal{D}$ requires at least $\Omega(2^{n/3-1})$ queries. 

Let us now turn to $\D_\Cl$. For any $s\in\{0,1\}^n$, the distribution $P_s$ is the output distribution of the quantum circuit shown in Fig.~\ref{fig:parity_with_noise}, without the red box, and with the $\mathrm{CNOT}$ gates determined by the string $s$. 
Moreover, any Clifford unitary $U\in\mathrm{Cl}(2^n)$ can be realized by a depth $d=O(n)$ circuit consisting only of nearest neighbour two-qubit Clifford gates~\cite{Bravyi_2021}.
It therefore follows that the output distributions of linear depth local Clifford circuits are exponentially hard to learn from statistical queries. \newline

\noindent Using \Cref{lem:embedding} we will now trade the query complexity for the depth at which the hardness sets in. In the previous paragraph, we have shown that learning depth $d$ Clifford distributions from statistical queries with tolerance $\tau(n)=\Omega(2^{-n/3})$ requires at least $q(n)=\Omega(2^{n/3})$ queries for any depth $d(n)=\Omega(n)$.
Let $n\leq g(n)= o(2^n)$, $d=\Omega(n)$ and define $d'(n)=d\circ g^{-1}(n)$.
Thus $\omega(\log(n))=g^{-1}(n)\leq n$ and $\omega(\log(n))=d'(n)$.
\Cref{lem:embedding} then implies that learning $\D_\Cl(n,d'(n))$ from statistical queries with tolerance $\tau'(n)=\tau\circ g^{-1}(n)$ requires at least  $q'(n)=q\circ g^{-1}(n)$ many queries with
\begin{align}
    &\Omega(2^{-n/3})=\tau'(n)=2^{-\omega(\log(n))}\\
    &q'(n)=2^{\omega(\log(n))}\,.
\end{align}
In particular, for any $\eps,\delta<1/2$ any statistical query algorithm for learning super logarithmic depth Clifford circuit distributions with inverse polynomial tolerance requires super polynomially many queries.  \\

\noindent To obtain the second claim we first apply Corollary  \ref{cor:solovay-kitaev} in conjunction with Lemma \ref{lem:stat-query-approximation-reduction} to find that the statistical query complexity for learning $\D_\G(n, d\log^c(n\cdot d/\sigma))$ with tolerance $\tau>\sigma$ is lower bounded by that of learning $\D_\Cl(n,d)$ with any tolerance $\gamma<\tau-\sigma$.
Now fix any $\tau\in\Omega(1/\poly(n))$, let $\gamma=\tau/3$ and $\sigma=\tau/3$ such that $\gamma+\sigma<\tau$. 
As just shown, for any $\eps,\delta<1/2$ we know that $(\eps,\delta)$-learning $\D_\Cl(n,\omega(\log(n)))$ requires $\omega(\poly(n))$ statistical queries with tolerance at least $\gamma=\Omega(1/\poly(n))$. 
Hence, for any $\eps<1/2-\sigma$ and $\delta<1/2$, we find that $(\eps,\delta)$-learning $\D_\G(n, d)$ takes $\omega(\poly(n))$ many statistical queries of tolerance $\tau$ with
\begin{align}
    d=\omega\lr{\log(n)\cdot\log^c\lr{\frac{n\cdot\log(n)}{\tau}}}
    =\omega\lr{\log^{c+1}(n)}
    \,,
\end{align}
where 
we have used $1<\tau^{-1}$.
Setting $k=c+1$ completes the proof. Importantly, as noted earlier, for some gate sets $c=1$ and hence $k=2$ \cite{harrow_efficient_2002}.
\end{proof}

\end{document}